\documentclass[a4paper,fleqn,usenatbib]{mnras}

\usepackage{newtxtext,newtxmath}

\usepackage[T1]{fontenc}
\usepackage{ae,aecompl}

\usepackage{graphicx}	
\usepackage{amsmath}	

\defcitealias{Hirashita:2020aa}{HM20}
\defcitealias{Hirashita:2019aa}{HA19}
\defcitealias{Mathis:1977aa}{MRN}

\title[PAHs and grain size distribution]{Evolution of grain size distribution with enhanced
abundance of small carbonaceous grains in galactic environments}

\author[H. Hirashita]{
Hiroyuki Hirashita$^{1,2}$\thanks{E-mail: hirashita@asiaa.sinica.edu.tw}
\\
$^{1}$Institute of Astronomy and Astrophysics, Academia Sinica,
Astronomy-Mathematics Building, No.\ 1, Section 4,
Roosevelt Road, Taipei 10617, Taiwan\\
$^{2}$Theoretical Astrophysics, Department of Earth and Space Science,
Osaka University, 1-1 Machikaneyama, Toyonaka, Osaka 560-0043, Japan
}

\date{Accepted XXX. Received YYY; in original form ZZZ}

\pubyear{2022}

\begin{document}
\label{firstpage}
\pagerange{\pageref{firstpage}--\pageref{lastpage}}
\maketitle

\begin{abstract}
We propose an updated dust evolution model that focuses on the grain size distribution
in a galaxy. We treat the galaxy as a one-zone object and include five main processes
(stellar dust production, dust destruction in supernova shocks, grain growth by accretion
and coagulation, and grain disruption by shattering). In this paper,
we improve the predictions related to small carbonaceous grains, which are responsible for
the 2175 \AA\ bump in the extinction curve and the polycyclic aromatic hydrocarbon
(PAH) emission features in the dust
emission spectral energy distribution (SED), both of which were underpredicted in
our previous model. In the new model, we hypothesize that
small carbonaceous grains are not involved in interstellar processing.
This avoids small carbonaceous grains being
lost by coagulation. We find that this hypothetical model shows a much better match to
the Milky Way (MW) extinction curve and dust emission SED than the previous one.
The following two additional modifications further make the fit to the MW dust emission SED
better:
(i) The chemical enrichment model is adjusted to give a nearly solar metallicity in the present
epoch, and the fraction of metals available for dust growth is limited to half.
(ii) Aromatization for small carbonaceous grains is efficient, so that the aromatic
fraction is unity at grain radii $\lesssim 20$ \AA. As a consequence of our modelling,
we succeed in obtaining a dust evolution model that explains the MW extinction curve and
dust emission SED at the same time.
\end{abstract}


\section{Introduction}

Dust grains in the interstellar medium (ISM) absorb and scatter
ultraviolet (UV) and optical light and reprocess
it in the infrared (IR).
The spectral energy distribution (SED) of a galaxy from UV to IR wavelengths
is thus actively shaped by dust extinction (absorption + scattering) and
reemission, depending on the dust opacity as a function of wavelength
\citep[e.g.][]{da-Cunha:2008aa,Boquien:2019aa,Abdurrouf:2021aa}.
Thus, it is important to clarify how dust interacts with radiation at various wavelengths.

The absorption and scattering cross-sections of dust depend on
the grain size distribution (distribution function of grain radius)
and composition {\citep[e.g.][]{Desert:1990aa,Li:1997aa,Weingartner:2001aa}}.
Some models, which aimed at constraining the grain size distribution and
dust composition, have been developed to account for the observed extinction curve
(extinction as a function of wavelength) and dust emission SED of the Milky Way
\citep[MW; e.g.][]{Dwek:1997aa,Draine:2001aa,Li:2001aa,Zubko:2004aa,Compiegne:2011aa,Jones:2017aa}.
These models prepare a set of dust materials (e.g.\ silicate, graphite, and amorphous carbon),
and choose the grain size distributions that fit the observed extinction curve and/or the
dust emission SED.
The models are also applied to nearby galaxies in order to extract the information on dust properties
{\citep[e.g.][]{Li:2002aa,Draine:2007aa,Galliano:2018aa,Relano:2020aa,Relano:2022aa}}.
All these models are useful not only to understand dust extinction and emission but also to infer
the dust composition and grain size distribution.

Although the dust properties in the MW and nearby galaxies have been revealed by the
above methods, it is yet to be clarified
if the obtained grain size distributions and grain compositions are explained as a result of
dust evolution in galaxies. It should be useful to clarify what kind of information on dust evolution
can be extracted from the observed dust properties or the grain size distributions and compositions
derived from observations.
Indeed, \citet{ODonnell:1997aa} made a pioneering effort of building up an
evolution model of grain size distribution for the purpose of explaining the observed MW extinction
curve.

There have been some theoretical models for the evolution of grain size distribution in a galaxy.
\citet{Asano:2013aa} incorporated various processes that affect the grain size distribution into a
single framework, and showed that the grain size distribution changes drastically as the galaxy evolves. According to their model together with later modified versions \citep{Nozawa:2015aa,Hirashita:2019aa},
the evolution of grain size distribution occurs in the following way.
In the early epoch, the dust abundance is dominated by stellar dust production from
supernovae (SNe) and asymptotic giant branch (AGB) stars. As the ISM is
enriched with metals and dust, the interplay between shattering and accretion (dust growth
by the accretion of gas-phase metals) plays a significant role in
increasing small grains. In a later epoch when the metallicity is nearly solar, coagulation
efficiently converts small grains to large ones, so that the grain size distribution approaches a shape
determined by the balance between coagulation and shattering. This balance leads to a grain
size distribution similar to the one derived by \citet[][hereafter MRN]{Mathis:1977aa}, which reproduces
the MW extinction curve.

The extinction curve and dust emission SED are sensitive not only to the grain size distribution
but also to the grain compositions. In particular, some dust species can be identified through
some features in the MW extinction curve and SED.
The prominent bump in the MW extinction curve
at $\lambda=2175$~\AA\ ($\lambda$ is the wavelength) can be explained by carbonaceous species.
Small graphite grains \citep{Stecher:1965aa,Gilra:1971aa} and
polycyclic aromatic hydrocarbons \citep[PAHs;][]{Mathis:1994aa} are
candidates for the bump carriers \citep[see also][]{Li:2001aa,Weingartner:2001aa,Steglich:2010aa}.
\citet{Jones:2013aa} assumed that the 2175~\AA\ bump originates from
aromatic-rich amorphous carbon.
All the above studies attribute the 2175 \AA\ bump to carbon materials with organized atomic structures
like aromatic carbons and graphite.
Aromatic species also contribute to prominent mid-IR (MIR)
emission features, whose carriers are considered to be  PAHs
\citep{Leger:1984aa,Allamandola:1985aa,Tielens:2008aa,Li:2012aa}
or some form of aromatic species {\citep{Jones:2013aa,Yang:2016aa,Yang:2017aa}},
although other variations
such as hydrogenated amorphous carbons \citep{Duley:1993aa},
quenched carbonaceous composite \citep{Sakata:1984aa}, and
mixed aromatic/aliphatic organic nanoparticles \citep{Kwok:2011aa} may also be responsible for the
features.

In our previous evolution model of grain size distribution \citepalias{Hirashita:2020aa},
the carriers of the 2175 \AA\ bump and the PAH features\footnote{We refer to
the prominent MIR features attributed to PAHs to the PAH features, although
the carriers are debated as mentioned above.} are assumed to be
small aromatic  carbonaceous grains.
In this model, the carbonaceous materials are
classified into two categories: `aromatic' and `non-aromatic', corresponding to
regular and irregular (amorphous) atomic structures.
We considered aromatization and
aliphatization for the change between the aromatic and non-aromatic species following
\citet{Murga:2019aa}.
We calculated the extinction curve and dust emission SED
by assuming that small aromatic grains are the carriers of
the 2175 \AA\ bump and the PAH features (\citetalias{Hirashita:2020aa}; \citealt{Hirashita:2020ad}).

The above model explains the correlation between metallicity and PAH feature strength
observed in nearby galaxies \citep[e.g.][]{Engelbracht:2005aa,Draine:2007ab,Galliano:2008aa,Ciesla:2014aa}
and in distant galaxies \citep[e.g.][]{Shivaei:2017aa}.
Since small carbonaceous grains are predominantly formed by interstellar processing
(shattering, etc.) of preexisting grains in the above model, the efficiency of their formation
depends on the dust abundance, predicting
a positive relation between
the PAH abundance and the dust-to-gas ratio, which correlates with the metallicity
\citep{Seok:2014aa,Rau:2019aa}.
Although interstellar processing is not the only way of explaining the metallicity dependence of
PAH abundance
\citep[e.g.\ contribution from AGB stars:][]{Galliano:2008aa,Bekki:2013aa},
and radiation hardness may play a role in regulating the PAH abundance
\citep[e.g.][]{Madden:2006aa,Hunt:2010aa,Khramtsova:2013aa},
our model that includes interstellar processing still provides a viable tool for
testing evolution scenarios of dust and PAHs in galaxies.

In spite of the broad success in our previous dust evolution model, there are problems
if we look closely into the abundance of small carbonaceous grains.
As argued by \citet{Weingartner:2001aa} and \citet{Li:2001aa}, the grain size distribution
has an excess at small grain radii for carbonaceous dust. This excess is particularly needed
to explain the strengths of PAH emission features in the MW. Such an excess in PAH abundance
was hard to reproduce in our model, which predicts a smooth power-law-like 
(\citetalias{Mathis:1977aa}-like) shape of grain size distribution as a result of the balance
between coagulation and shattering
in an evolved (metal/dust-enriched) ISM. Indeed, we observe in \citetalias{Hirashita:2020aa} that
our model underpredicts the 2175 \AA\ bump strength (see their fig.\ 3a) and
in \citet{Hirashita:2020ad} that PAH emission is
underestimated if we adopt a
model that explains the SEDs at $\lambda >20~\micron$ in the MW
(see their fig.\ 1b).
Since these features in the extinction curve and SED are important in observationally
characterizing the dust properties, it is worth investigating if we could remedy the
underestimating tendency of small carbonaceous grains in our dust processing scenario.

Some hints for how to improve the models for the carriers of
the 2175 \AA\ bump and PAH features can be
obtained from some previous results in the literature. We itemize them in what follows.
\begin{enumerate}
\item As mentioned above, some fitting models for the dust--PAH emission SED of the MW
introduce an excess
of small aromatic grains in the grain size distribution. \citet{Weingartner:2001aa} and
\citet{Li:2001aa} assumed the excess size distributions of PAHs in the form of
lognormal functions, and \citet{Jones:2013aa}
adopted a steeply rising power-law size distribution towards small grain radii. They commonly
indicated a significant excess of small carbonaceous grains compared
with the \citetalias{Mathis:1977aa} grain size distribution.
The deviation from the \citetalias{Mathis:1977aa}-like grain size  distribution implies that
small carbonaceous (or aromatic) grains should not be processed by shattering or coagulation,
since the balance between these two processes inevitably lead to
an \citetalias{Mathis:1977aa}-like grain size distribution as mentioned above
\citep[see also e.g.][]{Dohnanyi:1969aa,Tanaka:1996aa,Kobayashi:2010aa}.
\item Coagulation likely produces grain porosity \citep{Ormel:2009aa,Okuzumi:2009aa}.
For small carbonaceous grains, if we use graphite optical constants, the peak of the bump
shifts as the porosity develops as mentioned by \citet[][section 2.4]{Hirashita:2021aa}. This shift is inconsistent with the stability of the central
wavelength of the bump \citep[e.g.][]{Fitzpatrick:2007aa}.
This implies that, if the 2175~\AA\ carrier is graphite (or a material with similar optical
constants), it should not be processed by coagulation.
Extinction curve modelling by \citet{Voshchinnikov:2006aa} also adopted a
compact graphite component to fit the bump although they adopted porous grains for
silicate.
\end{enumerate}
The above two points seem to imply that the carriers of the 2175 \AA\ bump and the MIR features,
if both of them are small carbonaceous grains with regular atomic structures (aromatic carbon
or graphite), are not efficiently involved in interstellar processing
(especially coagulation) that usual dust grains experience.
This may arise from
different chemical properties (or chemical stability) of small carbonaceous grains.
Thus, an attractive modification of our model would be
to suppress the processing of small carbonaceous grains in the ISM.

In this paper, thus, we modify our evolution model of grain size distribution by
`turning off' interstellar processing for small carbonaceous grains.
Although this suppression of interstellar processing is just a hypothesis (without firm confirmation
by experiments or theory), it is worth investigating if this hypothesis leads to a quantitative match to
the 2175 \AA\ bump and PAH features observed in the MW.
This approach is also taken as an extreme case in which small carbonaceous grains
including PAHs are treated separately from other grain components.
Note again that we do not
strictly distinguish aromatic carbons and graphite for the simplicity of our
modeling since both of them have regular atomic
structures. Indeed, the 2175 \AA\ bump could also be explained by PAHs \citep{Li:2001aa};
thus, it is rather convenient not to distinguish these two species.
We refer to this hypothetical model as the \textit{new} model,
and distinguish it from the \textit{old} model in our previous development
(\citetalias{Hirashita:2020aa}).

This paper is organized as follows.
In Section~\ref{sec:model}, we describe our model for the evolution of grain size distribution
and the calculations of extinction curve and dust emission SED.
In Section~\ref{sec:result}, we show the basic results, which are compared with the MW data.
In Section~\ref{sec:discussion}, we further make an effort of making the model consistent with
the observed MW properties.
We also discuss the
robustness of the results and possible future improvements.
In Section \ref{sec:conclusion}, we give our conclusions.

\section{Model}\label{sec:model}

The model adopted to calculate the evolution of grain size distribution is based
on a one-zone galaxy evolution model developed by
\citetalias{Hirashita:2020aa} (old model).
In the new model, we modify the old model for the
treatment of small aromatic grains.
Based on the calculated grain size distributions of various dust species, we compute
the extinction curve and the dust emission SED.

\subsection{Review of the evolution model of grain size distribution}\label{subsec:review}

We review the model used to calculate the evolution of grain size distribution.
The model is based on \citetalias{Hirashita:2020aa}, and the full equations for dust
evolution processes are given by \citet{Asano:2013aa} and \citet{Hirashita:2019aa}.
We only give the outline below and refer the interested reader to these three papers
for further details.

The grain size distribution is denoted as $n_i(a)$
($i$ distinguishes the dust species and $a$ is the grain radius):
$n_i(a)\,\mathrm{d}a$ is the number density of grains whose radii are
between $a$ and $a+\mathrm{d}a$. We assume the grains to be spherical,
so that the grain mass, $m$, is related to $a$
as $m=(4\upi /3)a^3s$, where $s$ is the grain material density.
The grain size distribution is considered in the range of $a=3$ \AA--10 $\micron$
with 128 logarithmic bins,
and grains going out of the boundaries are removed by imposing the boundary
condition $n_i=0$ at the minimum and maximum grain radii.

The metallicity, which is calculated based on a chemical evolution model,
is used to trace the stellar dust production.
We adopt an exponentially decaying SFR with a time-scale of $\tau_\mathrm{SF}$
and  the Chabrier initial mass function \citep{Chabrier:2003aa} 
with a stellar mass range of 0.1--100 M$_{\sun}$.
Since we are only interested in MW-like evolution, we adopt $\tau_\mathrm{SF}=5$ Gyr
following \citetalias{Hirashita:2020aa} for the fiducial model.
The chemical evolution model calculates the metallicity ($Z$) as well as
the mass abundances of silicon and carbon ($Z_\mathrm{Si}$ and $Z_\mathrm{C}$, respctively),
which are used to determine the fractions of silicate and carbonaceous dust later.
{Since the \citetalias{Hirashita:2020aa} model was developed for the purpose of
describing generic dust evolution in galaxies, the output metal abundances have not
been calibrated with the MW abundances in detail. For the continuity of our studies,
we still show the output based on \citetalias{Hirashita:2020aa} with the minimum modifications
described below, but later make some efforts of
obtaining consistent metallicity and dust abundance with the MW observations.}

We calculate the dust enrichment by stars (SNe and AGB stars) based on
the increment of the metallicity, assuming that 10 per cent of newly ejected metals are
condensed into dust.
The produced dust by stellar sources is
distributed in each grain radius bin following the lognormal grain size distribution
centred at $a=0.1~\micron$ with a standard deviation of 0.47.
We also calculate the SN rate, which is used to
evaluate the dust destruction rate, in a manner consistent with the
star formation history.

For the evolution of grain size distribution, we consider interstellar processing of dust:
dust destruction by SN shocks, dust growth by the accretion of
gas-phase metals in the dense ISM, grain growth (sticking) by coagulation
in the dense ISM and grain fragmentation/disruption by
shattering in the diffuse ISM.
We consider the above processes that have been included in our
previous modelling, and neglect other processes such as rotational disruption by radiative torques
\citep{Hoang:2019aa,Hoang:2019ab}.

For simplicity, we fix the mass fraction of the dense ISM,
$\eta_\mathrm{dense}$ (we choose the fiducial value 0.5 in this paper), and adopt
$(n_\mathrm{H}/\mathrm{cm}^{-3},\, T_\mathrm{gas}/\mathrm{K})=(0.3,\, 10^4)$
and $(300,\, 25)$ for the diffuse and dense ISM, respectively,
where $n_\mathrm{H}$ is the hydrogen number density and $T_\mathrm{gas}$ is the gas temperature.
{These physical conditions are similar to those in the warm neutral (or ionized) medium
and the molecular clouds, where shattering and coagulation predominantly occur, respectively
\citep{Yan:2004aa}.}
The evolution of $n_i$ through coagulation and shattering is treated by
the Smolukovski equation (or its modification) with weighting factors of $\eta_\mathrm{dense}$ and
$(1-\eta_\mathrm{dense})$, respectively, while that through accretion and SN destruction
is treated by the `advection' equation in the $a$ space with weights of $\eta_\mathrm{dense}$ and unity,
respectively (unity means that the process occurs in both phases).

\subsection{New treatment for the small carbonaceous grains}\label{subsec:small_carbon}

As argued in the Introduction, we develop a hypothetical model in which
small carbonaceous grains are not affected by interstellar processing.
More precisely, we hypothesize that small carbonaceous grains,
{once they are formed mainly by shattering,} remain unprocessed by
coagulation, accretion, and shattering.
After some experimental runs, we found that turning off coagulation is the most
essential in raising the abundance of small carbonaceous grains.
This is because coagulation drives the
grain size distribution towards the \citetalias{Mathis:1977aa} shape, smoothing out any
excess grain size distribution at small radii.
Turning off shattering in addition to coagulation also helps keeping small
carbonaceous grains (in our treatment, grains shattered below the
lower grain radius limit, $a=3$ \AA, are lost).
We also turn off accretion. {High temperatures achieved by stochastic heating
\citep{Draine:2001aa} may also suppress accretion.}
We still include
{destruction in the hot gas associated with SN shocks} for small carbonaceous grains
since, otherwise, the dust abundance exceeds the metal abundance.
{Small carbonaceous grains are indeed destroyed by the collisions with
energetic ions and electrons in shocks
\citep[e.g.][]{Micelotta:2010aa}. We emphasize, though, that accretion and supernova destruction}
are less essential since coagulation and shattering
have a larger influence on the functional shape of grain size distribution at 
ages appropriate for the MW.

The grain radius below which carbonaceous dust is decoupled from interstellar processing
(except SN destruction)
is denoted as $a_\mathrm{C,dec}$.
We define the grain size distribution of decoupled carbonaceous species,
$n_\mathrm{C,dec}(a)$, as
\begin{align}
n_\mathrm{C,dec}(a)=
\begin{cases}
n_\mathrm{C}(a)\left[ 1-(a/a_\mathrm{C,dec})^3\right] & \text{if $a\leq a_\mathrm{C,dec}$,}\\
0 & \text{if $a>a_\mathrm{C,dec}$.}
\end{cases}
\end{align}
In this way, we empirically (or experimentally) realize a steep but continuous ransition from
unprocessed small carbonaceous
grains to processed large carbonaceous grains around $a=a_\mathrm{C,dec}$.
In calculating coagulation, shattering, and accretion for carbonaceous dust,
we use the grain size distribution, $n_\mathrm{C}'(a)=n_\mathrm{C}(a)-n_\mathrm{C,dec}(a)$
not to process the decoupled portion of the carbonaceous species.
After these processes, we add the unprocessed small carbonaceous grains to restore the
total carbonaceous grain size distribution as
$n_\mathrm{C}(a)=n'_\mathrm{C}(a)+n_\mathrm{C,dec}(a)$ at each time-step.

We adopt $a_\mathrm{C,dec}=20$ \AA.
If we adopt a smaller value of $a_\mathrm{C,dec}$, the difference between the old and new
models is too small to obtain significant improvement. If we adopt a larger value,
the 2175 \AA\ bump proves to become too strong. Since $a_\mathrm{C,dec}$ is
a completely free parameter in our model, we choose its value (20 \AA)
to optimize the improvement in the new model.

The above procedure has a risk of obtaining a dust-to-metal ratio higher than
unity after adding the unprocessed carbonaceous grain size distribution, even though
we minimize this risk by including SN destruction.
Thus, we impose an upper limit for the dust-to-metal ratio, $(\mathrm{D/M})_\mathrm{max}$,
and skip accretion if the dust-to-metal ratio exceeds this value. With a sufficiently small
time-step, this practically keep the dust-to-metal ratio below $(\mathrm{D/M})_\mathrm{max}$.
Unless otherwise stated, we adopt $(\mathrm{D/M})_\mathrm{max}=0.99$; that is,
almost all the metals can potentially condense into dust {as assumed by our original model
\citepalias{Hirashita:2020aa}. Indeed, as we will show later, this assumption leads to too high
a dust-to-metal ratio for the MW environment. Thus, we later}
examine a lower value of
$(\mathrm{D/M})_\mathrm{max}$ in Section \ref{subsec:less_growth} because
not all metals may be readily accreted onto dust.

\subsection{Fraction of each dust species}

We calculate the evolution of grain size distribution for silicate and carbonaceous dust
separately by assuming that \textit{all} the grains are composed of a single species
(silicate or carbonaceous dust).
The grain size distributions calculated in this way are denoted as $n_\mathrm{sil}(a)$ and
$n_\mathrm{C}(a)$ for silicate and carbonaceous dust, respectively.
We later average these two grain size distributions with appropriate weights.
In this method, we consider the different material properties
(difference in accretion time, tensile strength for disruption, etc.; see \citealt{Hirashita:2019aa})
are reflected in the grain size distribution, but we still avoid complication arising from
compound species.
The total grain size distribution, $n_\mathrm{tot}(a)$, is evaluated by the following weighted
average:
\begin{align}
n_\mathrm{tot}(a)=f_\mathrm{sil}n_\mathrm{sil}(a)+(1-f_\mathrm{sil})n_\mathrm{C}(a),
\end{align}
where $f_\mathrm{sil}=6Z_\mathrm{Si}/(6Z_\mathrm{Si}+Z_\mathrm{C})$ is the silicate
fraction (the abundance of silicon is multiplied by 6 to account for the fraction of silicon
in silicate grains; \citetalias{Hirashita:2020aa}).

In fact, the functional shape of the grain size distribution is not significantly different
between the two specie
at ages of interest for the MW, because the grain size distribution robustly
converges to an \citetalias{Mathis:1977aa}-like shape by the balance between shattering
and coagulation.
We refer the interested reader to  \citetalias{Hirashita:2020aa} for the difference
in shattering, coagulation, and accretion efficiencies between silicate and carbonaceous dust.
Note that the overall level of grain size distribution is
inversely proportional to $s$ ($s=3.5$ and 2.24 g cm$^{-3}$ for
silicate and carbonaceous dust, respectively; \citealt{Weingartner:2001aa})
if the total dust abundance is the same.

To predict aromatic features, we decompose the carbonaceous grain size distribution
into aromatic and non-aromatic species with the aromatic fraction $f_\mathrm{ar}(a)$.
In particular, small aromatic carbonaceous grains are regarded as PAHs.
We calculate $f_\mathrm{ar}(a)$ using the rate equations describing the exchange
between aromatic and non-aromatic species through
aromatization and aliphatization
occurring predominantly in the diffuse and dense ISM, respectively
\citep[the rates given in \citetalias{Hirashita:2020aa} are based on][see also \citealt{Jones:2012aa}]{Murga:2019aa}.
We neglect photo-processes that could act locally around massive stars
such as photo-destruction
\citep[e.g.][]{Andrews:2015aa,Chastenet:2019aa}
because they are difficult to include in our one-zone treatment
{and would be reasonablly examined in future spatially resolved studies.
As mentioned in \citetalias{Hirashita:2020aa}, the diffuse radiation with
$U=1$ (10), where $U$ is the interstellar radiation field normalized to the
solar neighbourhood value,  does not destroy PAHs with $a<3$ (5) \AA\ in 10 Gyr.
Thus, we neglect
the PAH destruction by the diffuse UV radiation and leave the destruction by locally
strong radiation field for future spatially resolved studies.}
The effect of the mean interstellar radiation field is included in the rate of aromatization,
{in which UV processing plays an important role as also shown by an experimental study
\citep{Duley:2015aa}}.
Based on the obtained $f_\mathrm{ar}(a)$, we decompose the carbonaceous grain size
distribution into aromatic and non-aromatic components as
\begin{align}
n_\mathrm{ar}(a)=f_\mathrm{ar}(a)n_\mathrm{C}(a),~~~
n_\text{non-ar}(a)=[1-f_\mathrm{ar}(a)]n_\mathrm{C}(a),
\end{align}
where $n_\text{ar}(a)$ and $n_\text{non-ar}(a)$ are the grain size distributions of
aromatic and non-aromatic species, respectively.

Here we clarify the output grain size distributions of the \textit{old} and \textit{new} models.
In both models, we calculate $n_\mathrm{sil}(a)$ and $n_\mathrm{C}(a)$ separately.
In the new model, we apply the new treatment of small carbonaceous grains
described in Section \ref{subsec:small_carbon}. In the old model, we do not apply
this new treatment; that is, we take into account interstellar processing also for
small carbonaceous grains following \citetalias{Hirashita:2020aa}.
Therefore, $n_\mathrm{C}(a)$ is different between the new and old models.
Since there is no modification for silicate,
$n_\mathrm{sil}(a)$ is identical between the new and old models.

\subsection{Extinction curves}\label{subsec:ext}

We first calculate the extinction curves separately for the silicate, aromatic, and non-aromatic
components using $n_\mathrm{sil}(a)$, $n_\mathrm{ar}(a)$, and
$n_\text{non-ar}(a)$, respectively.
The extinction for the component $i$ ($i=\text{sil}$, ar, or non-ar)
at wavelength $\lambda $ in units of magnitude ($A_{i,\lambda }$)
is calculated as
\begin{align} 
A_{i,\lambda}=(2.5\log_{10} \mathrm{e})L\displaystyle\int_{0}^{\infty}
n_{i}(a)\,\upi a^{2}Q_{\mathrm{ext},i}(a, \lambda),
\end{align}
where $L$ is the path length (which is not necessary to specify in the end; see below),
and $Q_{\mathrm{ext},i}(a, \lambda)$ is the extinction cross-section normalized to
the geometric cross-section (evaluated
by using the Mie theory; \citealt{Bohren:1983aa}) for dust species $i$.
For the convenience of comparison, we adopt the same optical properties
as in \citetalias{Hirashita:2020aa}; that is,
we adopt astronomical silicate \citep{Weingartner:2001aa} for silicate,
graphite in the same paper for aromatic
aromatic carbonaceous grains, and
amorphous carbon (`ACAR')
taken from \citet{Zubko:1996aa} for non-aromatic grains
\citep[see also][]{Nozawa:2015aa,Hou:2016aa}.
The total extinction, $A_\lambda$, is calculated by
\begin{align}
A_\lambda =f_\mathrm{sil}A_\mathrm{sil,\lambda}+
(1-f_\mathrm{sil})(A_\mathrm{ar,\lambda}+A_\text{non-ar,$\lambda$}).\label{eq:total_ext}
\end{align}
We concentrate on the shape (not the absolute level) of extinction curve by normalizing the extinction to the
$V$ band ($\lambda =0.55~\micron$) value:
$A_\lambda /A_V$, so that $L$ is cancelled out.

{
It is also useful to examine $A_\lambda /N_\mathrm{H}$, where $N_\mathrm{H}$ is
the hydrogen column density ($n_\mathrm{H}L$). In this ratio, $L$ is calcelled out as well.
The above ratio, $A_\lambda /A_V$, reflects the functional shape of the grain size distribution
but the effect of dust abundance is cancelled out.
In contrast, the ratio $A_\lambda /N_\mathrm{H}$ is proportional to the dust-to-gas ratio.
In this paper, we focus on the shape of the extinction curve including the
height of the 2175 \AA\ bump and the steepness of the far-UV rise; thus, we adopt
$A_\lambda/A_V$.
The effect of dust abundance is tested by the dust emission SED. However, we later confirm in
Section \ref{subsec:ext_NH} that the models that reproduce the dust emission SED predict
the level of $A_\lambda /N_\mathrm{H}$ consistent with the MW data.
}

\subsection{SED model}\label{subsec:SED}

We adopt the framework developed by \citet{Draine:2001aa} and \citet{Li:2001aa} to
calculate the SED. Briefly,
the distribution function of temperature ($\mathrm{d}P_i/\mathrm{d}T$)
for each $a$ is calculated by considering the balance between
the heating by the interstellar radiation field and the cooling by IR radiation.
The SED (expressed by the intensity per hydrogen) of the grain component $i$
($i=\text{sil}$, ar, or non-ar)
is calculated by
\begin{align}
I_{i,\nu} (\lambda )=
\int_0^\infty\mathrm{d}a\frac{1}{n_\mathrm{H}}n_i(a)\upi a^2Q_{\mathrm{abs},i}(a,\,\nu )
\int_0^\infty\mathrm{d}T\, B_\nu (T)
\frac{\mathrm{d}P_i}{\mathrm{d}T},\hspace{-1cm}\nonumber\\
\end{align}
where $Q_{\mathrm{abs},i}(a,\,\nu)$ is the absorption cross-section normalized to the
geometric cross-section ($\nu$ is the frequency corresponding to $\lambda$),
and $B_\nu (T)$ is the Planck function.
For the dust heating,
we adopt the spectrum of the MW interstellar radiation field in the solar neighbourhood
from \citet{Mathis:1983aa}.
We compose the total intensity with an expression similar to
equation (\ref{eq:total_ext}) as
\begin{align}
I_\nu (\lambda) =f_\mathrm{sil}I_\mathrm{sil,\nu}(\lambda )+
(1-f_\mathrm{sil})[I_\mathrm{ar,\nu}(\lambda )+I_\text{non-ar,$\nu$}(\lambda )].
\end{align}

For the convenience of comparison, we adopt the same material properties needed to
calculate $Q_{\mathrm{abs},i}$ and $\mathrm{d}P_i/\mathrm{d}T$ as adopted by
\citet{Hirashita:2020ad}.
We take the dust properties from \citet{Draine:2007aa}. Among their dust species,
we apply astronomical silicate for the silicate component in our model.
For the aromatic component, we adopt their carbonaceous species,
which has a smooth transition from PAHs to graphite at $a\sim 50$~\AA\
\citep{Li:2001aa}.
We adopt the ionization fraction of PAHs as a function of grain radius following
\citet{Draine:2007aa}, originally
from \citet{Li:2001aa}, who calculated the ionization fractions
in various ISM phases and averaged them with appropriate weights.
For the non-aromatic component, we simply adopt graphite from \citet{Draine:1984aa},
i.e.\ \citet{Draine:2007aa}'s carbonaceous properties without PAH features,
for the convenience of using their framework of calculating $\mathrm{d}P_i/\mathrm{d}T$.
According to \citet{Hirashita:2014aa}, the difference in the mass absorption coefficient
between amorphous carbon taken from \citet{Zubko:1996aa}
and graphite is 40 per cent around the peak of the
IR SED ($\lambda\sim 150~\micron$).
Since the fraction of non-aromatic carbon in carbonaceous dust is $\sim 0.5$
(for $\eta_\mathrm{dense}=0.5$), the above 40 per cent uncertainty causes
$\sim 20$ per cent uncertainty in the far-IR (FIR; roughly from 60 to 300 $\micron$)
SED. This uncertainty is small enough to draw meaningful conclusions from
comparison with the observed MW SED.

\subsection{Observational data for comparison}\label{subsec:obs}

We use the data of the extinction curve and dust emission SED of the MW.
For the extinction curve, we adopt the data from
\citet{Pei:1992aa}.
For the IR SED, we use the data for the diffuse high Galactic latitude
medium from \citet{Compiegne:2011aa} as typical dust emission data
in the solar neighbourhood.
The data are taken by the Diffuse Infrared Background Experiment (DIRBE) instrument
on \textit{the Cosmic Background Explorer} (\textit{COBE})
and instruments onboard \textit{Herschel}.
The same data for the IR SED were also
used for comparison in our previous dust evolution models
\citep{Hirashita:2020ad,Chang:2022aa}.
In addition, since the above broad-band data do not really reflect the height of
PAH emission features, we also refer to the spectroscopic data taken by
the \textit{Infrared Space Observatory} (\textit{ISO}) shown by
\citet{Compiegne:2011aa}, originally from \citet{Flagey:2006aa}.

\section{Results}\label{sec:result}

\subsection{Grain size distribution}\label{subsec:size}

We present the calculated grain size distributions.
The difference between the old and new models is produced by the
different treatment of interstellar processing for carbonaceous grains
(Section \ref{subsec:small_carbon}).
In the early stage of evolution, when coagulation
is not important, the two models show similar results
(the evolution of grain size distribution is already shown in \citetalias{Hirashita:2020aa}).
In this paper, we concentrate on later ages, especially $t=10$ Gyr,
which is appropriate for the age of the MW.

\begin{figure}
\begin{center}
\includegraphics[width=0.48\textwidth]{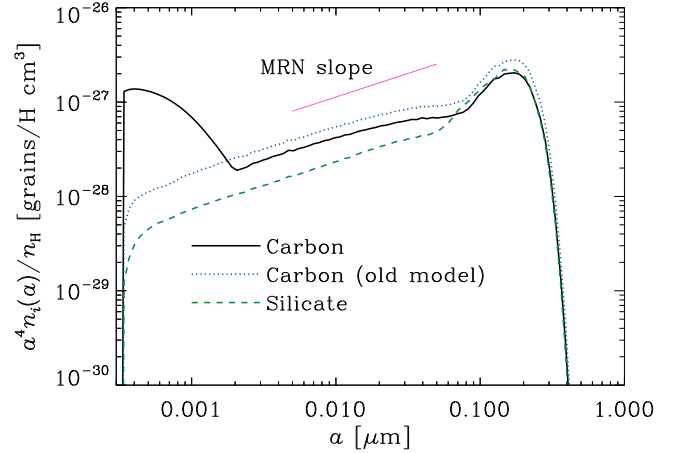}
\end{center}
\caption{Grain size distributions at $t=10$~Gyr,
multiplied by $a^4$ and divided by $n_\mathrm{H}$
(the resulting quantity is proportional to the mass weighted grain size distribution per
$\log a$ relative to the gas mass).
The solid and dashed lines correspond to the grain size distributions calculated with
the properties of carbonaceous dust
and silicate [i.e.\ $n_i(a)=n_\mathrm{C}(a)$ and $n_\mathrm{sil}(a)$], respectively,
in the new model.
The dotted line shows the grain size distribution in the old model using the carbonaceous
grain properties. Note that the silicate grain size distribution is identical between the new
and old models.
The thin straight line presents the slope of the MRN grain size distribution
($n_i\propto a^{-3.5}$).
\label{fig:size}}
\end{figure}

We show the calculated grain size distributions in Fig.~\ref{fig:size}.
We find that the grain size distributions computed with the carbonaceous
material propertie, $n_\mathrm{C}(a)$, has an excess at
$a<a_\mathrm{C,dec}$ ($=20$~\AA) in the new model. This is mainly
because coagulation is suppressed for small carbonaceous grains.
Suppression of shattering further keeps small carbonaceous grains from
being fragmented and lost from the lower boundary of the grain radius.
By comparing the new and old models, we observe that the grain size distribution of
carbonaceous dust at $a>a_\mathrm{C,dec}$ does not change significantly.
The slightly lower grain size distribution at $a>a_\mathrm{C,dec}$
in the new model is due to more grains being distributed at $a<a_\mathrm{C,dec}$.
The grain size distribution of silicate has a similar slope to that of carbonaceous dust in the old model,
and is approximately described by the \citetalias{Mathis:1977aa} distribution
($n\propto a^{-3.5}$). The different levels of grain size distributions between
the two species reflect the difference in gran material density (Section \ref{subsec:review}).

\subsection{Extinction curves}\label{subsec:ext_result}

\begin{figure}
\includegraphics[width=0.48\textwidth]{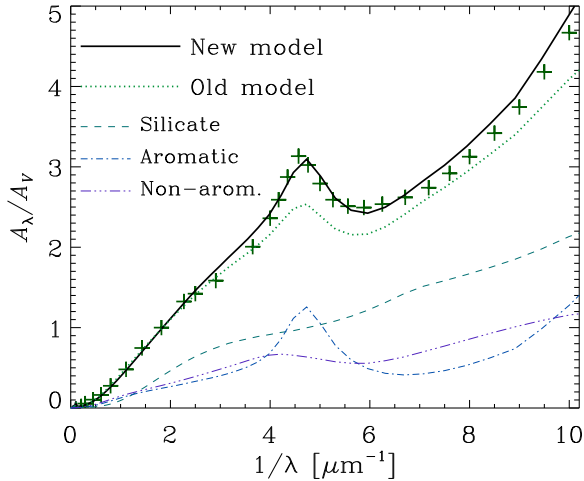}
\caption{Extinction curves normalized to the value in the $V$ band
at $t=10$ Gyr, computed from the grain size distributions shown in
Fig.\ \ref{fig:size} and the calculated fraction of each grain species.
The solid and dotted lines correspond to the new and old models,
respectively. The dashed, dot--dashed, and triple-dot--dashed lines
present the contribution from silicate, aromatic, and non-aromatic components,
respectively, in the new model.
The crosses show the observed MW extinction curve taken from
\citet{Pei:1992aa}.
\label{fig:ext}}
\end{figure}

Based on the grain size distribution and grain compositions in our
model at $t=10$~Gyr, we calculate the extinction curve (Section \ref{subsec:ext}).
At this age, $f_\mathrm{sil}=0.68$.
In Fig.~\ref{fig:ext}, we compare the extinction curves in the new and old models.
As mentioned in the Introduction, the old model underestimates
the 2175~\AA\ bump strength. The new model, as expected from the increase of
small aromatic grains (Fig.\ \ref{fig:size}), enhances the 2175 \AA\ bump,
predicting more consistent bump strength than the old model. Moreover,
the slope in the far-UV ($\lambda\lesssim 1500$ \AA)
is also better reproduced in the new model because
of more small grains. Thus, in our model, the excess of small carbonaceous
grains plays a significant role in reproducing the MW extinction curve.

We emphasize
that the fractions of the three grain species (silicate,
aromatic and non-aromatic components) are quantities that are
predicted from the model.
In our model, as shown in Fig.~\ref{fig:ext}, the contributions from silicate and carbonaceous
(aromatic + non-aromatic) dust
to the far-UV extinction are comparable. This is similar to the fitting results using a silicate--graphite
mixture by
\citetalias{Mathis:1977aa} and \citet{Weingartner:2001aa}.
{The comparable contribution from silicate and carbonaceous dust to the far-UV
extinction may be consistent with the lack of clear correlations between silicon and carbon depletions
and the far-UV rise \citep{Mishra:2015aa,Mishra:2017aa}.
and between the 2175 \AA\ bump strength and the far-UV extinction \citep{Xiang:2017aa}.}
The level of extinction for the aromatic and non-aromatic
components are also similar. The 2175~\AA\ bump is created by the aromatic component.
Therefore, the good match to the observed MW extinction
curve supports the fractions of the grain species.

\subsection{Dust emission SED}\label{subsec:SED_result}

\begin{figure}
\includegraphics[width=8cm]{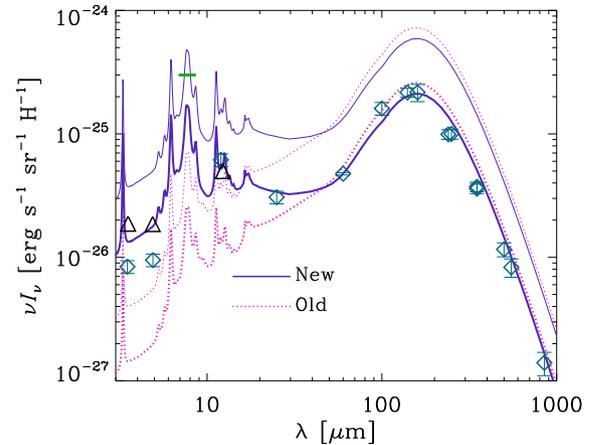}
\caption{Dust emission SEDs. The solid and dotted lines correspond to the
new and old models (at $t=10$ Gyr), respectively. The thin lines show the
output SEDs, while the thick lines show the SEDs divided by 2.8 to
better compare the SED shape with the observational data 
(square points with error bars) taken from \citet{Compiegne:2011aa}.
The triangles show the bandpass-averaged intensities of the model SED shown
by the thick solid line for
the DIRBE 3.5, 4.9, and 12 $\micron$ bands.
The horizontal green short line shows the peak level of the 7.7 $\micron$ feature
indicated by the spectroscopic data taken by \textit{ISO}
\citep{Flagey:2006aa,Compiegne:2011aa}.
\label{fig:sed}}
\end{figure}

Another observed characteristic that reflects the grain size distribution is the dust emission
SED, which is calculated based on the grain size distribution and grain compositions
at $t=10$~Gyr (Section~\ref{subsec:SED}).
In Fig.~\ref{fig:sed}, we show the {calculated} SEDs. As already shown in
\citet{Hirashita:2020ad}, our model overpredicts the SED in the FIR at $t=10$~Gyr.
Since the FIR emission reflects the total dust abundance, the overprediction is
due to an overabundance of dust. 
This overestimate of dust abundance arises from the high metallicity ($Z=0.021$) and
the high dust-to-metal
ratio, which almost reaches $\mathrm{(D/M)_{max}}=0.99$.

In our model, the contributions from various dust species (not shown for the
clarity of the figure) are
similar to those presented by \citet{Li:2001aa}, who also used
a mixture of silicate, graphite, and PAHs to fit the MW SED (see
fig.\ 8 of \citealt{Li:2001aa}).
The SED is dominated by the carbonaceous component.
Silicate only has a comparable intensity to carbonaceous dust at
$\lambda\gtrsim 300~\micron$.

It is still meaningful to discuss the SED \textit{shape} (SED renormalized to match the observed
FIR SED), which is affected by the functional shape of the grain size distribution.
We divide the calculated SED by 2.8 to make the level of FIR emission consistent with
the observation. As presented in
Fig.~\ref{fig:sed}, the new model
significantly enhances the emission at $\lambda\lesssim 20~\micron$ compared with the
old one. As a consequence,
the new model better reproduces the SED shape.
We also show the bandpass-averaged intensities for the DIRBE 3.3, 4.9, and 12~$\micron$
bands, where the PAH emission features are prominent. (At longer wavelengths,
the data points can be directly compared with the model lines because the bandpass-averaged
intensities are almost identical to the monochromatic intensity.)
At 3.3 and 4.9 $\micron$, the SED shape of the new model slightly overpredicts the emission,
while at 12 $\micron$, it matches the observation well.
The spectroscopic data are represented by the peak level of the 7.7 $\micron$ emission
not to complicate the figure, since the spectral profile is similar between the model and
observation \citep{Compiegne:2011aa}. The SED shape predicts two times lower
PAH emission features than the spectroscopic data.
Overall, the SED shape is reproduced by the new model
within a factor of $\sim$2.

\section{Improved models}\label{sec:discussion}

In the previous section, we showed that our new hypothetical prescription for the treatment of
small carbonaceous grains significantly improves the match to the observed MW
extinction curve and dust emission SED shape. However, we still found that
the absolute level of the FIR SED is significantly overpredicted,
indicating that the dust abundance is overestimated.
In fact, some parameters in the model are still uncertain so that it is worth varying
them. In this section, we mainly try to remedy the overprediction of dust abundance.
There are three possibilities for decreasing the
dust abundance: (i) If we adopt a younger age, the chemical enrichment proceeds less,
so that the dust abundance becomes lower. (ii) If we make the chemical enrichment slower,
the metallicity and dust abundance are lower at a fixed age.
(iii) If the efficiency of accretion (dust growth) is lower,
the dust-to-metal ratio decreases, which leads to a lower dust abundance.
We consider these three possibilities for better fitting to the
observed dust emission SED.
We also address the underprediction of PAH feature strength shown above in comparison with the
spectroscopic data.

\subsection{Younger age}\label{subsec:younger}

A younger age may be a solution for reducing the total dust abundance.
Since our model assumes a fixed star formation time-scale, the metallicity
is roughly proportional to the age \citep{Asano:2013aa}.
Thus, we examine the results at $t=3$~Gyr (3 times younger age)
to resolve the factor $\sim$3 overestimate of the FIR emission in
Section~\ref{subsec:SED_result}.
{We choose this age just to examine an earlier stage of dust enrichment,
and we do not aim at arguing that it better represents the MW age.}
At this age, $f_\mathrm{sil}=0.68$.

\begin{figure}
\includegraphics[width=8cm]{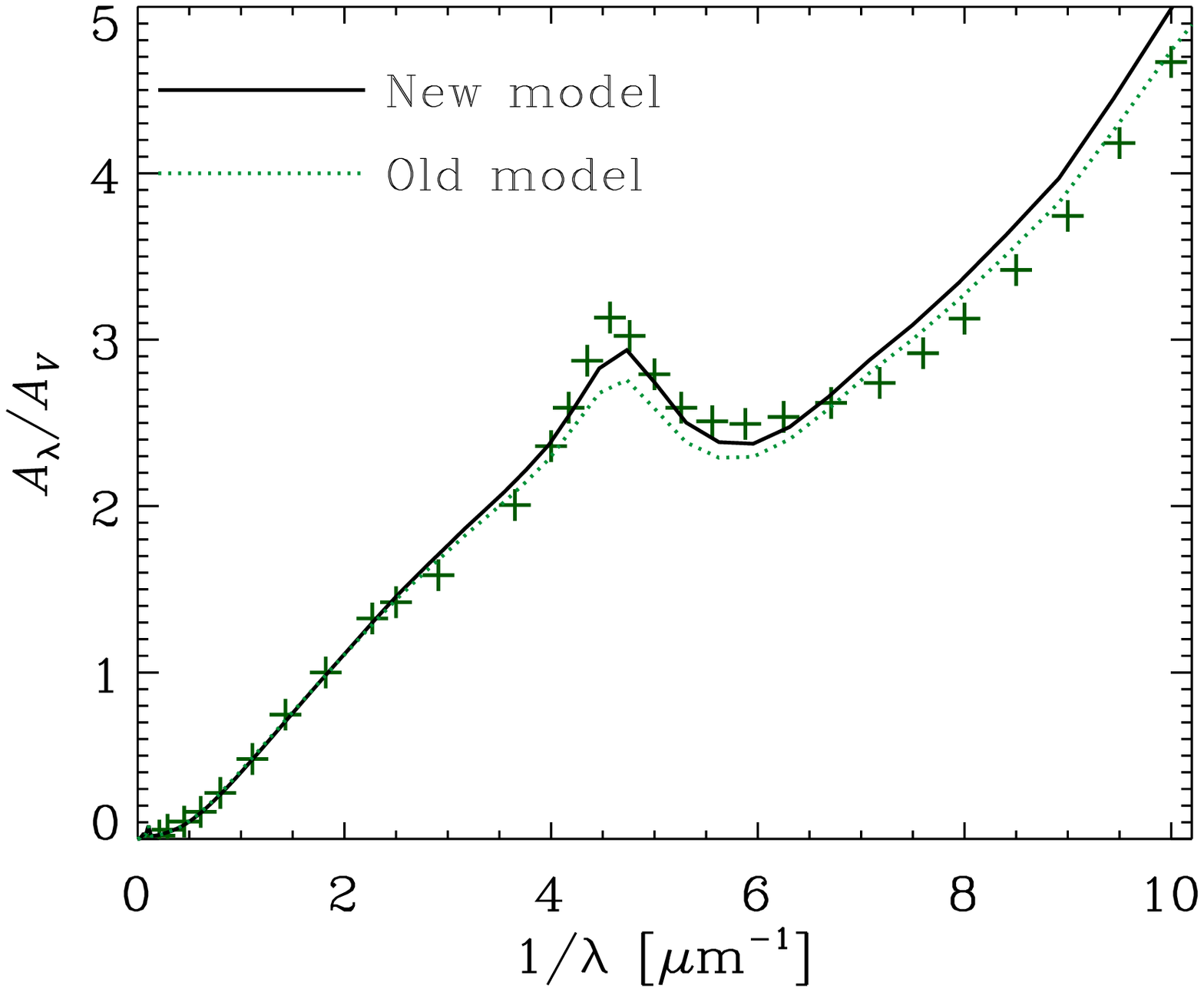}\\
\includegraphics[width=8cm]{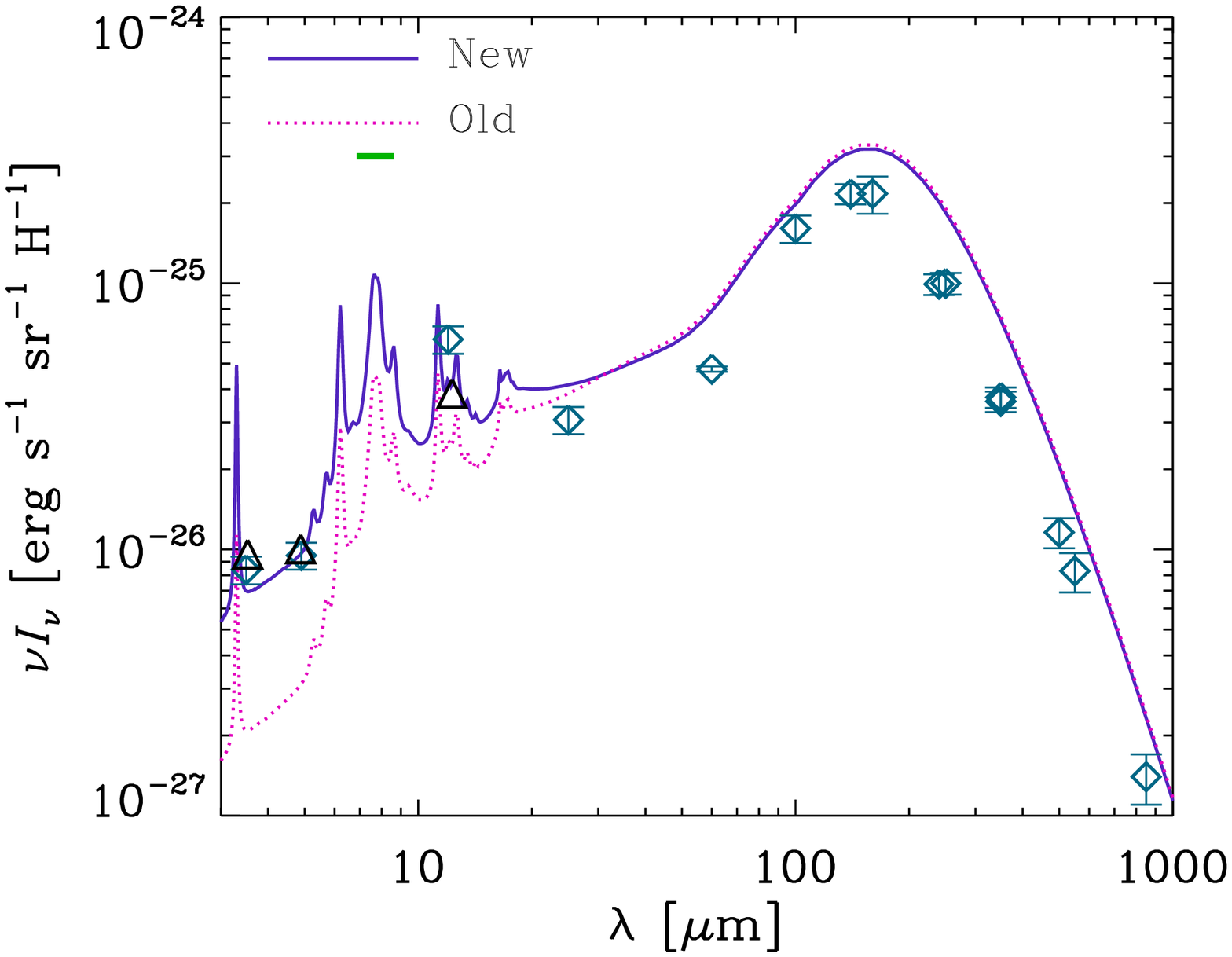}
\caption{Extinction curves (upper) and dust emission SEDs (lower panel)
at $t=3$ Gyr. The solid and dotted lines show the
new and old models, respectively. The same observational data as in
Figs.\ \ref{fig:ext} and \ref{fig:sed} are shown.
\label{fig:3Gyr}}
\end{figure}

In Fig.~\ref{fig:3Gyr}, we show the extinction curves and dust emission SEDs at
$t=3$~Gyr. For the extinction curves, the difference between the new and old models is
qualitatively similar to that at $t=10$ Gyr; that is, the new model predicts a stronger
2175 \AA\ bump and a steeper far-UV slope.
Although the new model reproduces the 2175 \AA\ bump,
it slightly overpredicts the far-UV slope.
Overall, the extinction curve is
steeper at $t=3$ Gyr than at $t=10$ Gyr because coagulation has not yet been efficient
enough for the grain size distribution to converge to an MRN-like shape
(i.e.\ a larger abundance of small grains than is expected from the
\citetalias{Mathis:1977aa} distribution;
\citetalias{Hirashita:2020aa}).

The IR SED presented in Fig.~\ref{fig:3Gyr} is indeed lower than that shown in
Fig.~\ref{fig:sed} (without factor 2.8 correction) because of lower dust abundance.
The model overproduces the MW SED by a factor of 1.5 at FIR wavelengths,
while it is consistent with the broadband data at short wavelengths. 
We still find that the calculated
SED underpredicts by a factor of 3 the PAH feature strength indicated by the spectroscopic data
(whose 7.7 $\micron$ feature peak is marked in the figure).

From the above, we conclude that a younger age ($\sim 3$ Gyr) could
provide a solution for the overprediction of the FIR emission.
However, the fit to the extinction curve is slightly worse and the PAH feature strength is
underestimated.
{We also confirmed that, if we choose ages less than 3 Gyr, the fit to the
extinction curve becomes even worse.}
Therefore, it is still worth examining the other two possibilities of reducing the
dust abundance; that is, the above items (ii) and (iii).

\subsection{Slower metal enrichment}\label{subsec:longer_tauSF}

Now we examine the second possibility: slower metal enrichment.
For this purpose, we adjust the star formation time-scale, $\tau_\mathrm{SF}$, which
directly regulates the metal enrichment time-scale in our model.
A $\tau_\mathrm{SF}$ much longer than the cosmic age
would predict too inefficient conversion from gas to stars.
Thus, we adopt $\tau_\mathrm{SF}=10$ Gyr (comparable to the cosmic age)
as the longest possible star formation time-scale for the MW.
This also makes the metallicity of the system comparable to that of the MW as
we show later.
The old model is also recalculated with $\tau_\mathrm{SF}=10$ Gyr.
In this scenario, $f_\mathrm{sil}=0.67$ at $t=10$ Gyr.

In this case, the resulting extinction curve and SED shape (not shown) are similar to those presented
above in Section \ref{sec:result} (Figs.\ \ref{fig:ext} and \ref{fig:sed}),
except for the normalization. This case overproduces the FIR SED by a factor of $\sim 1.8$.
This is because the dust-to-gas ratio is too high:
The metallicity $Z=0.015$ at $t=10$ Gyr is consistent with that in the MW
\citep{Asplund:2009aa}, but almost all
the metals are condensed in dust; i.e.\ the dust-to-metal ratio is nearly
$\mathrm{(D/M)_\mathrm{max}}=0.99$.
Thus, the overprediction is due to too high a dust-to-metal ratio.

The above result indicates that the
dust-to-metal ratio needs to be reduced to resolve the overestimating problem of
the MW FIR SED. Thus, the above solution (iii), that is, inefficient dust growth by accretion,
is worth investigating.

\subsection{Less efficient dust growth}\label{subsec:less_growth}

In our model, we assume that all the metals are available for dust growth.
However, as \citet{DeVis:2021aa} included in their model, it may be more realistic to
adopt an upper limit for the dust-to-metal ratio (i.e.\ for the fraction of metals that
can be condensed into dust), because
some metal elements are not easily or fully included in dust.
Considering that the above model with $\tau_\mathrm{SF}=10$ Gyr reproduces the
metallicity of the MW at $t=10$ Gyr,
we adopt the same model as in Section \ref{subsec:longer_tauSF}
(thus, $f_\mathrm{sil}=0.67$ is the same) but constrain the
dust-to-metal ratio. After some tests, we choose $\mathrm{(D/M)}_\mathrm{max}=0.48$
since this value reproduces the SED peak intensity.
The old model is also recalculated with the same condition for the metal enrichment
and dust growth.

\begin{figure}
\includegraphics[width=8cm]{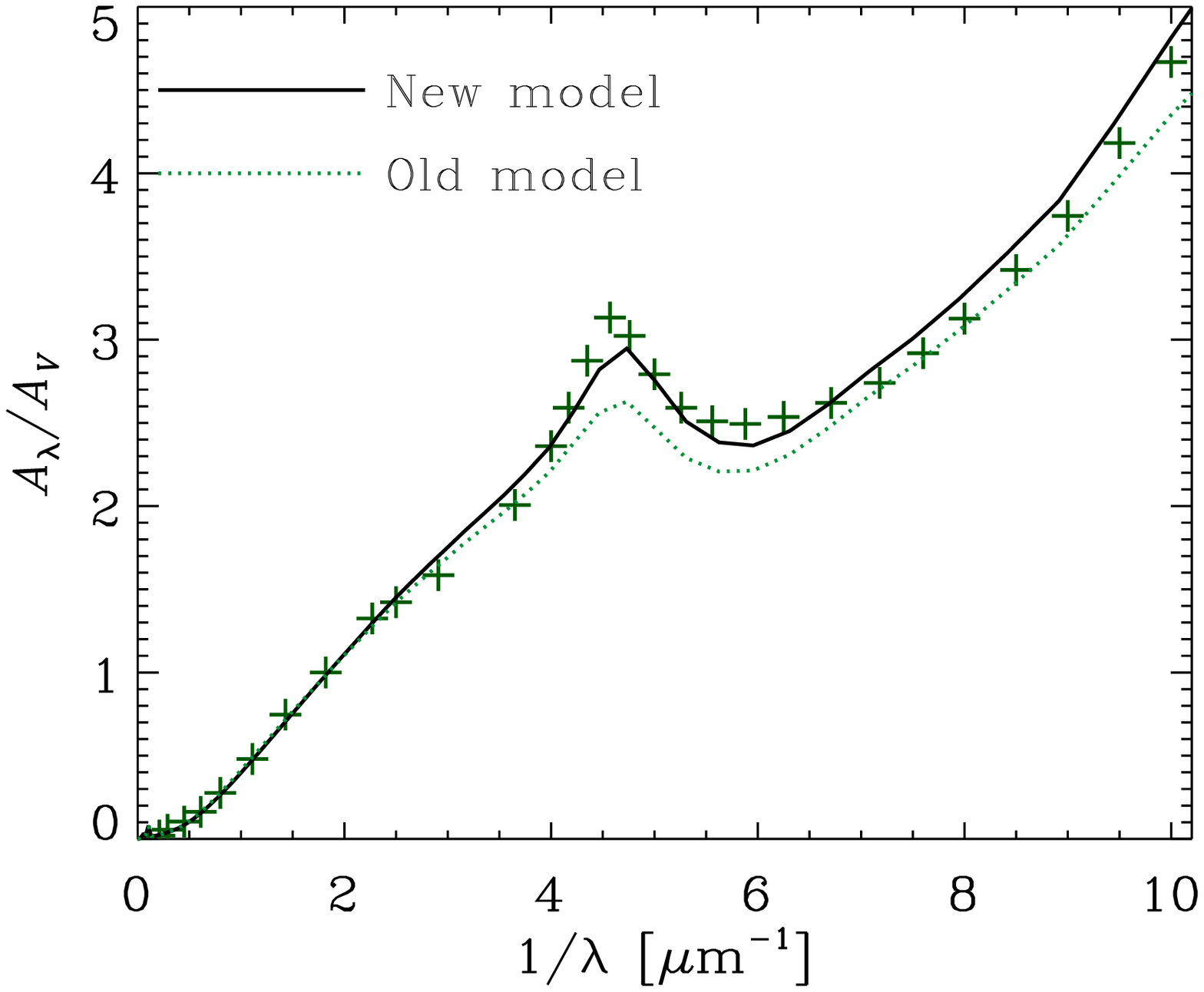}\\
\includegraphics[width=8cm]{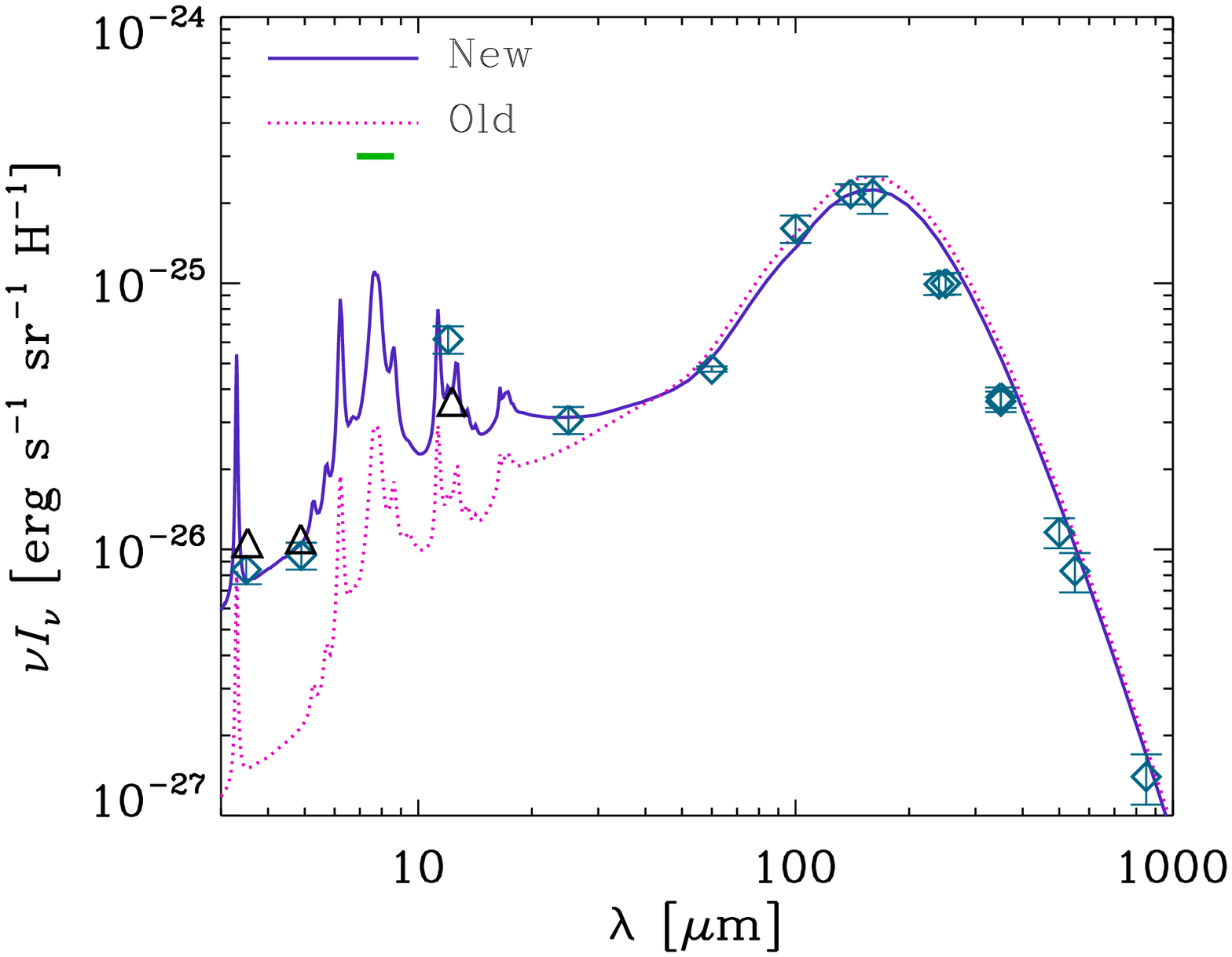}
\caption{Extinction curves (upper) and dust emission SEDs (lower panel)
at $t=10$ Gyr for the model investigated
in Section \ref{subsec:less_growth}, where we adopt
reduced efficiency of dust growth by accretion
(the maximum value of dust-to-metal ratio is assumed to be 0.48)
and slower metal enrichment with $\tau_\mathrm{SF}=10$ Gyr. The solid and dotted lines show the
new and old models, respectively. The same observational data as in
Figs.\ \ref{fig:ext} and \ref{fig:sed} are shown.
\label{fig:hybrid}}
\end{figure}

In Fig.~\ref{fig:hybrid}, we show the resulting extinction curves and dust emission SEDs.
We observe that the extinction curve fits the MW extinction curve well in the new model.
The old model underpredicts the 2175~\AA\ bump
strength. Thus, the new model improves the fit to the MW extinction curve significantly
also in this scenario.

We also show the dust emission SED in Fig.~\ref{fig:hybrid}.
Overall, the calculated SED fits the MW data points at all wavelengths
except for the 7.7~$\micron$ point from the spectroscopy.
The good match to the FIR SED is realized by construction because
we adjusted $\mathrm{(D/M)}_\mathrm{max}=0.48$ to reproduce the SED peak.
At short wavelengths, where PAH emission features are
prominent, the new model (the triangle points that take into account the
bandpass averaging) reproduce the observational data points
within a factor of $\sim$1.5, which is an excellent agreement with the data
considering that the old model significantly underestimated the emission.
The strength of the PAH features is still underpredicted, as is clear from the
comparison with the level of the 7.7 $\micron$ feature indicated by the spectroscopic data.

Now it is meaningful to discuss the abundances of silicate and carbonaceous dust
in this scenario with $\tau_\mathrm{SF}=10$ Gyr and 
$\mathrm{(D/M)}_\mathrm{max}=0.48$, because this model reproduces
the MW extinction curve and the overall level of the dust emission SED.
The total dust-to-gas ratio
is $7.0\times 10^{-3}$, which is consistent with often used values
\citep[e.g.][]{Weingartner:2001aa}. The dust-to-gas ratio almost reaches
the maximum value ($0.015\times 0.48=7.2\times 10^{-3}$) because of
dust growth by accretion. Since the silicate fraction is
$f_\mathrm{sil}=0.67$ in this model, the silicate dust-to-gas ratio is
$4.7\times 10^{-3}$ while the carbonaceous dust-to-gas ratio is
$2.3\times 10^{-3}$.
{\citet{Weingartner:2001aa} obtained constraints for the silicate and carbonaceous
dust-to-gas ratios as
$4.5\times 10^{-3}$ and $2.0\times 10^{-3}$, respectively, from the interstellar depletion.
These values are broadly consistent with
the above values.
The required abundances of the carbon and silicon atoms per hydrogen in number
for our dust model are
$2.6\times 10^{-4}$ and $3.9\times 10^{-5}$, respectively, which are below the
constraints for the present-day abundances, $3.4\times 10^{-4}$ and $4.3\times 10^{-5}$,
respectively
\citep[after taking the Galactic chemical evolution into account;][]{Zuo:2021aa,Zuo:2021ab,Hensley:2021aa}.}
If we define PAHs as aromatic grains whose radii are smaller than
12.8 \AA\ \citep[corresponding to the number of carbon atoms = $10^3$;][]{Draine:2007ab}, we obtain
the PAH-to-gas ratio ($q_\mathrm{PAH}$) as 2.6 per cent.
This is smaller than the value of $q_\mathrm{PAH}$ recommended by \citet{Draine:2007ab}
($q_\mathrm{PAH}=4.6$ per cent) for the MW. The smaller value of $q_\mathrm{PAH}=2.6$ per cent
in our model explains the underprediction of the PAH features.

\subsection{Maximum PAH prescription}\label{subsec:maxPAH}

The above underpredictions of PAH feature strength are worth investigating further,
since the aromatic fraction in the model can be uncertain.
The aromatic fraction is roughly $f_\mathrm{ar}\simeq 1-\eta_\mathrm{dense}$
in our model since
aromatization is assumed to occur in the diffuse ISM.
However, if aromatization still occurs
in the dense ISM or if aliphatization does not occur efficiently, our method underestimates
the PAH fraction.
Some PAH emission is found to be associated with the dense ISM
\citep[molecular clouds; e.g.][]{Sandstrom:2010aa,Chastenet:2019aa}, which
implies that the aromatic fraction is kept high
at least for small grains whose radii correspond to PAH emission carriers
($a\lesssim 20$~\AA). Thus, we examine a possibility that
$f_\mathrm{ar}(a)=1$ at $a\leq 20$~\AA\ (we use the calculated values for $f_\mathrm{ar}$
at $a>20$ \AA) to investigate an extreme of efficient aromatization.
This is referred to as the maximum PAH prescription, and this prescription is applied to
both old and new models.
Other than this, we adopt the same setting as in Section \ref{subsec:less_growth},
with $\mathrm{(D/M)_{max}}=0.48$ and $\tau_\mathrm{SF}=10$ Gyr.

\begin{figure}
\includegraphics[width=8cm]{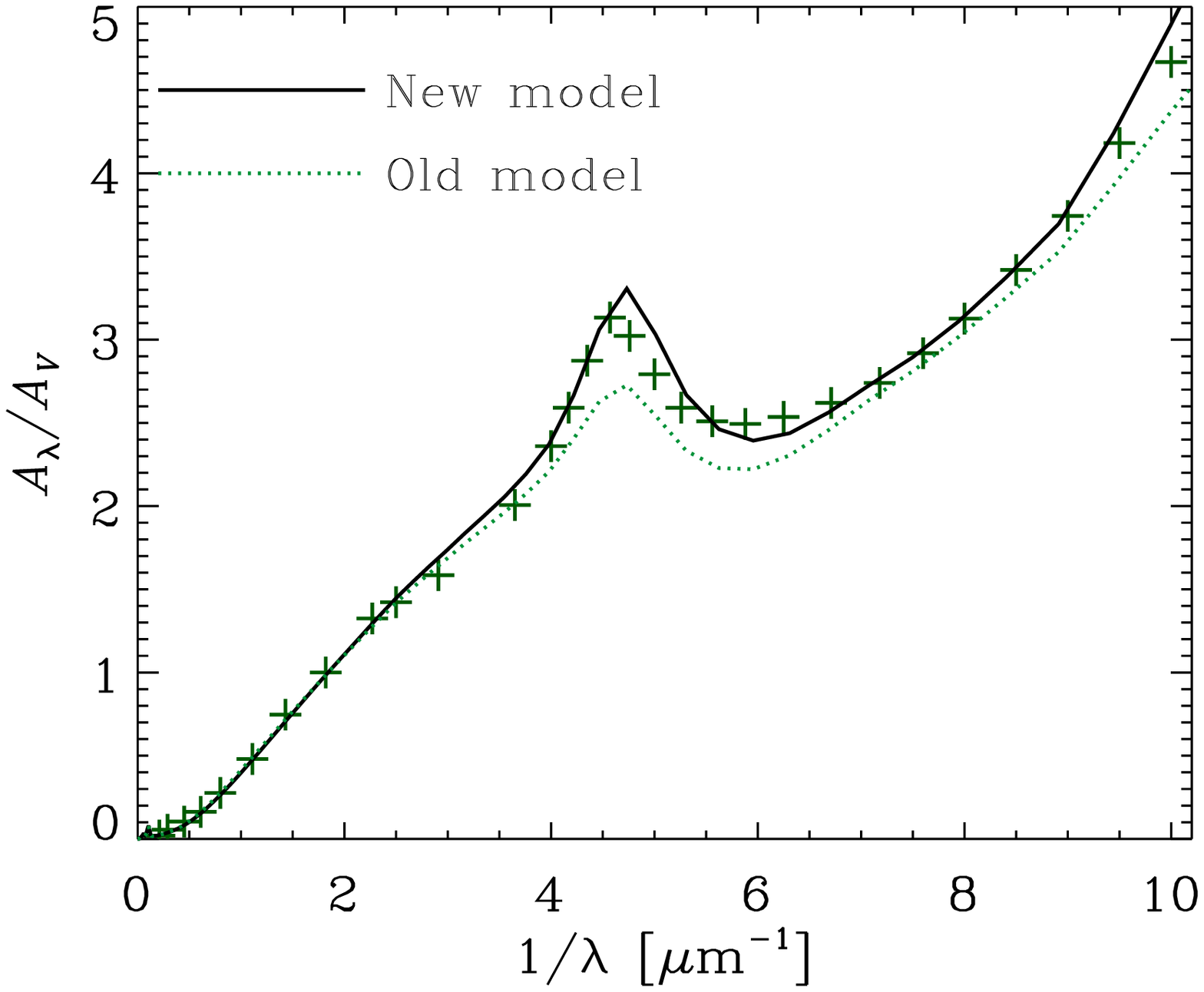}\\
\includegraphics[width=8cm]{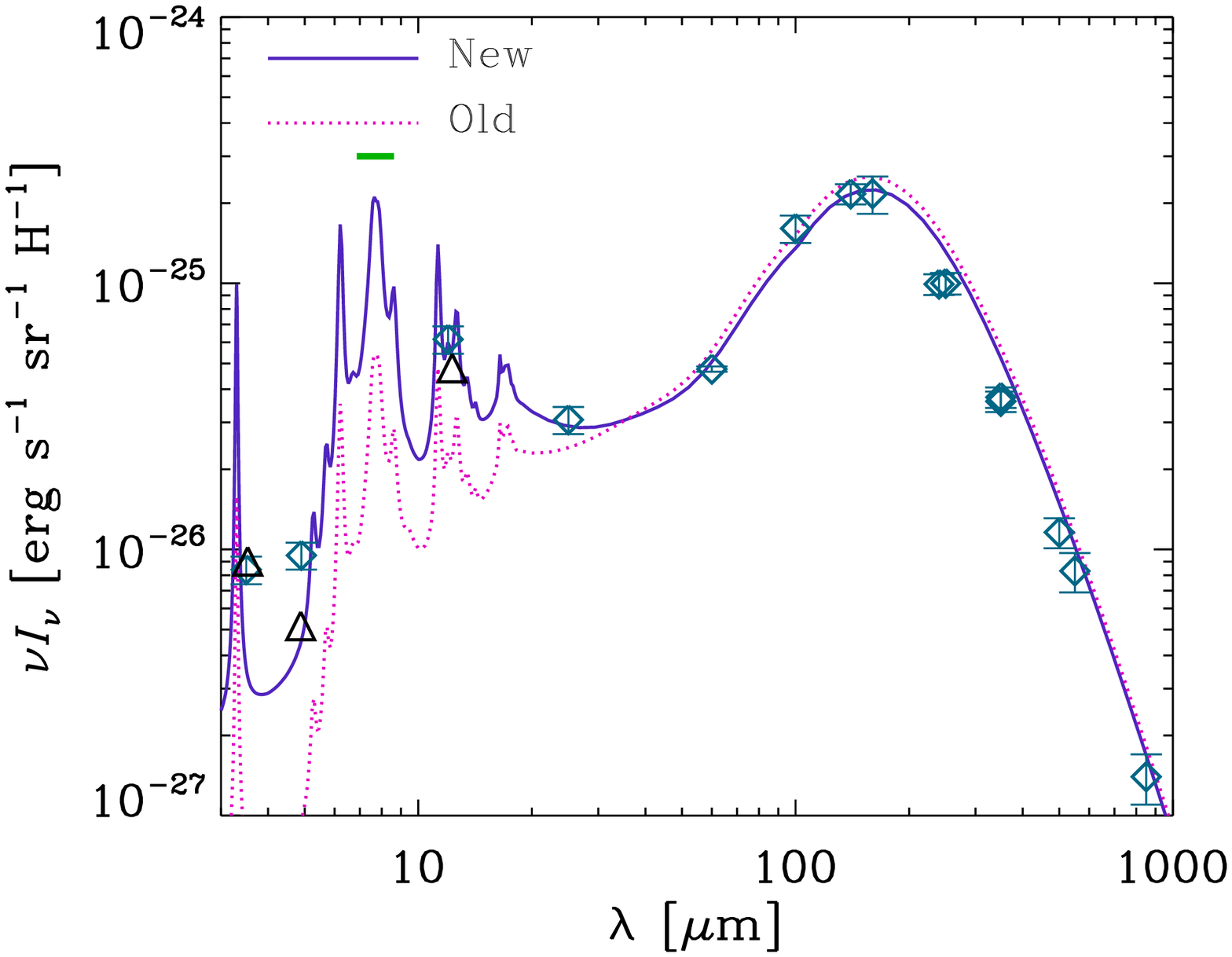}
\caption{Same as Fig.\ \ref{fig:hybrid} but for the maximum PAH prescription
in Section \ref{subsec:maxPAH}, where we maximize the aromatic fraction
at $a\leq 20$ \AA\ with $\mathrm{(D/M)_{max}}=0.48$ and $\tau_\mathrm{SF}=10$ Gyr.
\label{fig:maxPAH}}
\end{figure}

In Fig.~\ref{fig:maxPAH}, we show the extinction curves and dust emission SEDs
with the maximum PAH prescription.
If we concentrate on the new model, which is significantly better than the old model,
we find that the fit to the extinction curve is still good
(even slightly better than in Fig.~\ref{fig:hybrid}). Therefore, the change of the aromatic fraction
at $a\leq 20$~\AA\ does not affect (or worsen) the good fit to the extinction curve.
Also, we still observe a good fit to the SED data points at $\lambda >20~\micron$.
The SED data points at wavelengths where PAH emission is prominent
are also broadly good; in particular, the peak of the 7.7~$\micron$ feature is near to the
level indicated by the spectroscopic data. The 3.3 and 12~$\micron$ data show a better
match, while fit to the 4.9 $\micron$ point becomes worse because of the lower continuum
level. Since this continuum level is uncertain (highly dependent on the graphite mixture in
the cross-section formula in equation 2 of \citealt{Li:2001aa}), we regard the fit as being
satisfactory.

In this model, $q_\mathrm{PAH}$ is doubled compared with the value in
Section \ref{subsec:less_growth}; that is, $q_\mathrm{PAH}\sim 5.2$ per cent,
which is near to the value ($q_\mathrm{PAH}=4.6$ per cent)
adopted by \citet{Draine:2007ab}. Thus, our model is able to explain the PAH fraction as high
as $\sim 5$ per cent if we assume efficient aromatization for small carbonaceous grains.

\subsection{Absolute level of extinction}\label{subsec:ext_NH}

{Here we check if the level of extinction is actually reproduced or not.
As mentioned in Section \ref{subsec:ext}, we focused on the extinction curve shape
represented by $A_\lambda /A_V$ to test the functional shape of the grain size distribution
rather than the total dust abundance.
The dust abundance level was tested by the SED.
However, it is still useful to check if the absolute strength of extinction is
reproduced or not. To this goal, we examine $A_\lambda /N_\mathrm{H}$, which
reflects the dust abundance relative to hydrogen as well as the grain size distribution.}

\begin{figure}
\includegraphics[width=8cm]{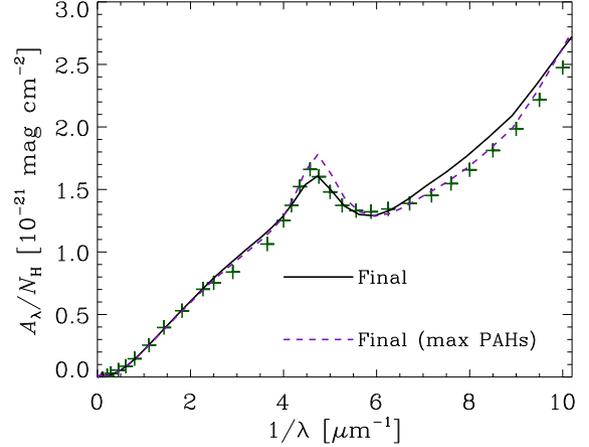}
\caption{{Extinction curves per hydrogen for the final two models in Sections \ref{subsec:less_growth}
and \ref{subsec:maxPAH} (solid and dashed lines, respectively). For comparison, we show the
observed MW extinction curve by crosses taken from \citet{Pei:1992aa} with the normalization
given by \citet{Weingartner:2001aa}.}
\label{fig:ext_NH}}
\end{figure}

{In Fig.\ \ref{fig:ext_NH}, we show the extinction curve per hydrogen for the
models in Sections \ref{subsec:less_growth} and \ref{subsec:maxPAH}, which broadly
reproduced the dust emission SED.
For comparison, we show the MW extinction curve taken from \citet{Pei:1992aa}
and adopt the normalization $A_V/N_\mathrm{H}=5.3\times 10^{-22}$ mag cm$^2$
\citep{Weingartner:2001aa}.
From the figure, we
confirm that these models also reproduce the level of extinction per hydrogen.
We also confirmed (not shown) that
other cases that overpredicted the dust emission
SED (i.e.\ models in Sections \ref{subsec:SED_result}, \ref{subsec:younger}, and \ref{subsec:longer_tauSF})
also overestimate the extinction by a similar factor. Therefore, we prove the
consistency between the levels of
dust extinction and emission per hydrogen in our model.}

\subsection{Possible improvements}\label{subsec:improvement}


The optical properties of dust may need further studies.
When we calculated extinction curves in our models, we adopted graphite properties
for the aromatic component for the purpose of direct comparison with and continuity from
our previous models
(\citetalias{Hirashita:2020aa}; \citealt{Hirashita:2020ad}).
This can be modified
by including PAH optical properties as done by
\citet{Li:2001aa}; that is, we apply a continuous transition from
PAH to graphite for the aromatic component as the grain radius becomes larger.
The transition occurs around $a\sim 50$~\AA.
Using this PAH--graphite model for the aromatic component in our model in Section \ref{subsec:maxPAH},
we find (not shown) that the 2175~\AA\ bump and far-UV slope are overproduced.
This is because the enhanced
contribution from PAHs raises both the 2175 \AA\ bump and the far-UV slope.
Thus, as far as the UV extinction curve is concerned, using the graphite optical properties,
which are not too sensitive to the grains at $a\lesssim 20$ \AA,
leads to a better fit to the extinction curve in our model.


Different sets of optical properties are also worth trying.
There have recently been significant updates in the models of dust optical properties
by \citet[][see also \citealt{Draine:2021ab}]{Hensley:2022aa}.
In their new models, the silicate and graphite components are unified to `astrodust',
but PAHs are still included as a distinct component.
An enhanced grain size distribution at $a\lesssim 20$ \AA\ is still needed for the PAH
component; thus, our new model is still useful to explain such an enhancement.
In the future work, it would be
interesting to use their new models for the optical properties of dust.
There are some other models for dust optical properties. We discuss
two often used models from \citet{Jones:2013aa} and \citet{Zubko:2004aa}.
\citet{Jones:2013aa} adopted (hydrogenated) amorphous carbon, including aromatic
materials,
and amorphous silicate to explain the MW dust properties
\citep[see also][and references therein]{Jones:2017aa}.
Their model assumes a grain size distribution of small carbonaceous grains steeply
rising towards small grain radii.
\citet{Zubko:2004aa}'s models also require an excess abundance
of PAHs relative to the extension of the \citetalias{Mathis:1977aa} grain size distribution
towards small grain radii.
Thus, it seems that we generally require a
large enhancement of small carbonaceous grains to explain the observed
dust properties in the MW, especially the strong PAH emission features.

The following two extensions, which need more detailed dust models and/or
different frameworks, will be necessary. First,
comparisons with other observational quantities such as polarization,
near-IR extinction curves \citep[e.g.][]{Hensley:2021aa}, would be useful.
Secondly, it is also interesting to treat hydrodynamic
evolution of the ISM, which our one-zone treatment is not able to include.
For hydrodynamic evolution, previous simulations in which
the evolution of grain size distribution is implemented
\citep{McKinnon:2018aa,Aoyama:2020aa,Li:2021ab,Romano:2022ab,Romano:2022aa}
can be extended to include the evolution of PAHs.

Finally, we should keep in mind that the decoupling of small carbonaceous
grains from interstellar processing (coagulation, accretion, and shattering)
is still a hypothesis. This decoupling probably arises from the chemical properties 
(e.g.\ stability) of
small carbonaceous species, but we still need to clarify detailed physical and chemical processes.

\section{Conclusions}\label{sec:conclusion}

In this paper, we make an effort of improving the treatment of small carbonaceous
grains to explain the MW extinction curve and dust emission SED at the same time.
We use the evolution model of grain size distribution developed in our previous studies
(\citealt{Hirashita:2019aa}; \citetalias{Hirashita:2020aa}). This model
treats the galaxy as a one-zone object, and includes
stellar dust production, dust destruction by SNe, dust growth by accretion,
grain growth by coagulation, and grain disruption by shattering.
The silicate-to-carbonaceous dust mass ratio is determined from the abundance ratio
between Si and C, while the aromatic fraction is estimated from the balance between
aromatization and aliphatization.
Small aromatic grains are responsible for the 2175 \AA\ bump in extinction and
the PAH emission features.
In the old model,
the 2175 \AA\ bump and PAH emission were significantly underestimated.
Thus, we develop a new hypothetical model in which we assume that
small carbonaceous grains (including PAHs)
are not processed by coagulation, shattering and accretion (probably because of
chemical stability). This modification particularly suppresses the loss of
small carbonaceous grains by coagulation.
We also apply (to both old and new models)
a minor modification that treats silicate and carbonaceous dust
separately.

First, we adopt a
star formation time-scale $\tau_\mathrm{SF}=5$ Gyr, following the standard
choice in \citetalias{Hirashita:2020aa}.
We assume the age of the MW to be $t=10$ Gyr.
Since this model overpredicts the
metal abundance, we significantly overestimate the FIR dust emission.
However, if we divide the SED by 2.8 to focus on the SED shape,
it broadly reproduces the SED at all IR wavelengths except for the level of
the PAH emission features, which are significantly underpredicted.
The extinction curve
in the new model better reproduces the MW curve than in the old model
in terms of the 2175 \AA\ bump and the far-UV slope.

There are some ways to reduce the dust abundance to resolve the problem of
overestimating the FIR SED. One is to assume a younger age. If we adopt $t=3$ Gyr,
the extinction curve at this age is broadly consistent with the observed MW extinction
curve, but the far-UV slope is slightly overproduced.
The overall level of the SED is consistent with the broad-band data; however,
the PAH feature strength is still underpredicted, and the FIR emission is overpredicted.

We also examine a longer $\tau_\mathrm{SF}(=10~\mathrm{Gyr})$ to decrease the
metallicity, which could contribute to reducing the FIR emission.
At $t=10$~Gyr, the extinction curve and the SED shape (after renormalization to
match the SED peak) in this case are similar to
those in the case with $\tau_\mathrm{SF}=5$ Gyr. The FIR SED is still overestimated by a
factor of 1.8. Given that the metallicity ($Z=0.015$) is consistent with the MW value,
the overprediction of the FIR SED is due to the overestimate of the dust-to-metal ratio.
Too high a dust-to-metal ratio may arise
from our assumption that all metals are available for dust growth by accretion.

To reduce the accretion efficiency, we set an upper limit for the dust-to-metal ratio, beyond which
metals are not accreted onto the dust. This assumption reflects the fact that some elements
are not easily (or fully) condensed into dust. If we assume the upper limit for
the dust-to-metal ratio to be 0.48,
the FIR SED becomes consistent with the MW observations at $t=10$ Gyr.
At this age, the calculated extinction curve
fits the observed curve very well. The calculated dust emission SED fairly fits the broad-band
data of the
MW SED. However, the strength of
the PAH emission features is still underpredicted.

Finally, we propose a way to resolve the underestimate of the PAH emission features.
Since the treatment of aromatization may be uncertain, we examine a model in which
we maximize the PAH abundance by assuming the aromatic fraction to be unity at
$a\leq 20$~\AA. We find that this modification keeps the good fits to the MW extinction
curve and emission SED broad-band data. In addition, the observed strength of the PAH features
is nearly reproduced. Therefore, we finally
obtain a dust evolution model that fits both extinction curve and dust emission SED in the MW.

In summary, we have shown that the evolution model of grain size distribution, which
includes stellar dust production and interstellar processing, is made
consistent with the MW extinction curve and dust emission SED by applying
the following modifications: (i) Small ($a\lesssim 20$ \AA) carbonaceous grains
are decoupled from interstellar processing (coagulation, shattering, and accretion). This
causes an excess of small carbonaceous grains in the grain size distribution, and enhances
the 2175 \AA\ bump and PAH emission.
(ii) A combination of slower metal enrichment with $\tau_\mathrm{SF}=10$ Gyr
and more inefficient accretion with $\mathrm{(D/M)_{max}}=0.48$
reproduces the observed level of the MW FIR SED.
(iii) Efficient aromatization at $a\lesssim 20$ \AA\ is necessary. In particular, if all the
carbonaceous grains at $a\leq 20$ \AA\ are aromatic (i.e.\ PAHs),
the model is consistent with the MW extinction curve and dust emission SED.

The necessity of enhancing the PAH
abundance (as included in our new model) seems to be robust against different sets of dust
materials in the literature. Our hypothesis that PAHs are not involved in interstellar processing
(especially coagulation) provides a reason why the PAHs abundance is
kept high, although detailed physical processes that realize this hypothesis still need
clarifying.

We emphasize that the grain size distribution and the fraction of each grain
species are predicted
properties in our model. It is important that the grain size distribution
realized as a result of theoretically expected dust evolution is capable of
explaining the MW extinction curve and dust emission SED.
Our evolution model of grain size distribution provides an explanation for the
mechanisms that eventually manifest the actually observed extinction curve and dust emission
SED. We will address the robustness of our conclusion in future work by including
more complicated ISM evolution, physical and chemical properties of PAHs, etc.

\section*{Acknowledgements}
 
{We are grateful to the anonymous referee for useful comments.}
We thank the National Science and Technology Council for support through grants
108-2112-M-001-007-MY3 and 111-2112-M-001-038-MY3,
and the Academia Sinica for Investigator Award AS-IA-109-M02.

\section*{Data availability}
The data underlying this article will be shared on reasonable request to the corresponding author.



\bibliographystyle{mnras}
\bibliography{/Users/hirashita/bibdata/hirashita}


\appendix


\bsp	
\label{lastpage}
\end{document}